\documentclass[10pt,twocolumn,letterpaper]{article}

\usepackage{wacv}              % To produce the CAMERA-READY version
\definecolor{wacvblue}{rgb}{0.21,0.49,0.74}
\usepackage[pagebackref,breaklinks,colorlinks,allcolors=wacvblue]{hyperref}

\def\wacvPaperID{3379} % *** Enter the WACV Paper ID here
\def\confName{WACV}
\def\confYear{2026}

\usepackage{multirow}
\usepackage{colortbl}

\usepackage{soul}
\usepackage{makecell}
\usepackage{amsfonts}
\usepackage{mathtools}
\usepackage{booktabs}

\newcolumntype{g}{>{\columncolor{CuGray}}c}
\newcolumntype{z}{>{\columncolor{CuGray}}l}

\renewcommand{\paragraph}[1]{\noindent\textbf{#1.}\,\,}

\usepackage{xspace}

\def\onedot{.\@\xspace}
 
\def\ie{\emph{i.e}\onedot}

\newcommand{\Eref}[1]{Eq.~(\ref{#1})}
\newcommand{\Fref}[1]{Fig.~\ref{#1}}
\newcommand{\Tref}[1]{Table~\ref{#1}}

\newcommand{\be}{\begin{eqnarray}}
\newcommand{\ee}{\end{eqnarray}}
\newcommand{\bee}{\begin{eqnarray*}}
\newcommand{\eee}{\end{eqnarray*}}

\newcommand{\matrixb}{\left[ \begin{array}}
\newcommand{\matrixe}{\end{array} \right]}

\def\methodname{{\texttt{LighthouseGS}}}

\title{LighthouseGS: Indoor Structure-aware 3D Gaussian Splatting for Panorama-Style Mobile Captures}

\author{
Seungoh Han$^{1}$\thanks{Equal contribution}
\quad Jaehoon Jang$^{1}$\footnotemark[1]
\quad Hyunsu Kim$^{2}$
\quad Jaeheung Surh$^{2}$ \\
\quad Junhyung Kwak$^{2}$
\quad Hyowon Ha$^{2}\thanks{Corresponding author}$
\quad Kyungdon Joo$^{1}\footnotemark[2]$
\\
$^{1}$Ulsan National Institute of Science and Technology (UNIST) \quad $^{2}$Bucketplace \\
{\tt\small \{seungohhan00, wkdwogns1997, gustnxodjs, jaeheungsurh}\\
{\tt\small junhk0914, hyowonha.phd, kdjoo369\}@gmail.com}
}

\begin{document}
\maketitle
\begin{abstract}
We introduce \texttt{LighthouseGS}, a practical novel view synthesis framework based on 3D Gaussian Splatting that utilizes simple panorama-style captures from a single mobile device.
While convenient, this rotation-dominant motion and narrow baseline make accurate camera pose and 3D point estimation challenging, especially in textureless indoor scenes. 
To address these challenges, \texttt{LighthouseGS} leverages rough geometric priors, such as mobile device camera poses and monocular depth estimation, and utilizes indoor planar structures.
Specifically, we propose a new initialization method called plane scaffold assembly to generate consistent 3D points on these structures, followed by a stable pruning strategy to enhance geometry and optimization stability.
Additionally, we present geometric and photometric corrections to resolve inconsistencies from motion drift and auto-exposure in mobile devices. 
Tested on real and synthetic indoor scenes, \texttt{LighthouseGS} delivers photorealistic rendering, outperforming state-of-the-art methods and enabling applications like panoramic view synthesis and object placement.
Project page:~\href{https://vision3d-lab.github.io/lighthousegs/}{https://vision3d-lab.github.io/lighthousegs/}
\end{abstract}
\section{Introduction}
\label{sec:intro}
\textcolor{black}{Photorealistic novel view synthesis (NVS) for indoor scenes is essential for bringing real-world experiences into virtual worlds, enabling authentic interaction.
With the advent of neural rendering techniques, Neural Radiance Fields~\cite{mildenhall2020nerf} have demonstrated remarkable performance in NVS.
Recently, 3D Gaussian Splatting (3DGS)~\cite{kerbl20233d} has emerged as a powerful alternative, providing real-time rendering with high-fidelity view synthesis.
\textcolor{black}{Building on these approaches, several studies~\cite{huang2022real, yang2023nerfvs, xu2023VRNeRF, bai2024360} have explored their applications in AR/VR, such as indoor navigation and room tours.}
}
\textcolor{black}{However, despite the high-quality rendering of existing NVS methods, their practical deployment remains limited for general users (\ie, non-experts).
First, utilizing multi-camera rigs~\cite{kaien2020mdp, xu2023VRNeRF, turki2024hybridnerf} or 360$^\circ$ cameras~\cite{chen2023panogrf, bai2024360, li2025omnigs} allows for capturing the entire surrounding scene in a single shot, but these setups are often inaccessible due to their cost and complexity.
Second, even with a single mobile device, most prior works~\cite{kerbl20233d, scaffoldgs} still require hundreds of densely captured images with sufficient overlap.
This capturing process is also impractical for general users, who typically do not walk through an entire space to capture images from all angles.
}
\begin{figure}
    \centering
    \includegraphics[width=0.95\linewidth]{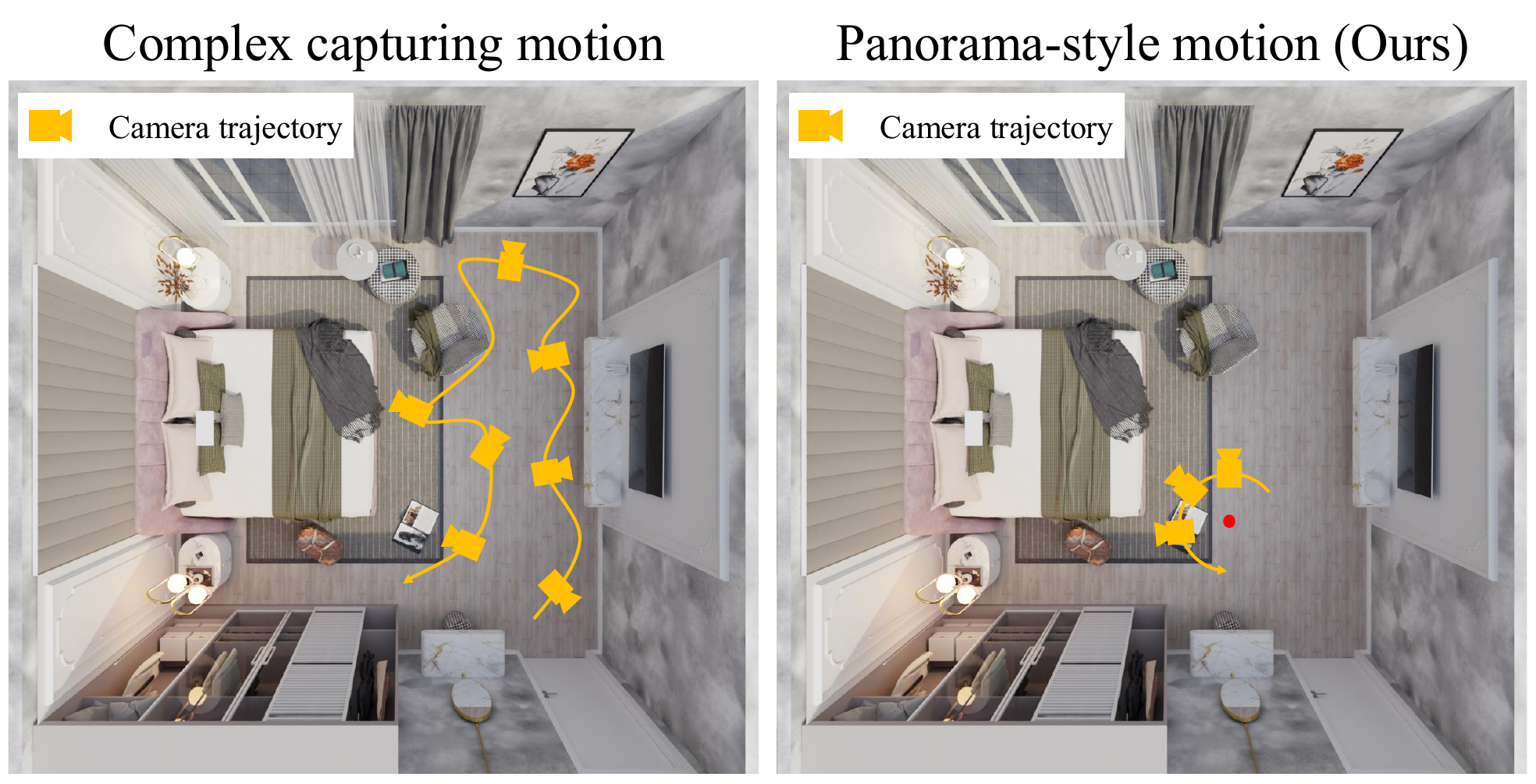}
    % \vspace{-1mm}
    \caption{
    {\textcolor{black}{Comparison between complex capturing motion and panorama-style motion for novel view synthesis}}.
    % \textit{Up}: 
    \textcolor{black}{}}
    \label{fig:overview}
\end{figure}
\textcolor{black}{Motivated by this fact, we explore a method for generating
photorealistic NVS from casually captured indoor scenes, aiming to improve accessibility for non-expert users.
Concretely, we leverage a \emph{panorama-style motion}~\cite{sweeney2019structure}, a natural and practical way for users to capture their surroundings with a single mobile device, as illustrated in \Fref{fig:overview}.
Panorama-style motion, where users stand in place and rotate the camera with half-stretched arms, enables efficient scene coverage without professional equipment or complex procedures.
However, integrating this motion into the NVS pipeline entails technical challenges.
Due to its rotation-dominant motion and narrow baseline, Structure-from-Motion (SfM)~\cite{schonberger2016structure} often fails to estimate accurate camera pose and reliable 3D points, which are crucial for initializing NVS frameworks.
These issues are further exacerbated in textureless regions of indoor scenes, resulting in degraded rendering quality.
}

\textcolor{black}{In this work, we propose a practical 3DGS-based NVS framework for panorama-style captures of indoor scenes by a mobile device camera.
Inspired by the panorama-style motion, which is similar to a lighthouse shining its light, we call the proposed method \texttt{LighthouseGS}.
To overcome the challenges of SfM under panorama-style motion and indoor scenes, we introduce a new initialization scheme, plane scaffold assembly, that exploits ARKit camera poses and monocular depth estimates.
Although these rough geometric priors may be imprecise due to IMU drift and scale ambiguity, plane scaffold assembly combines them with the planar structure of indoor scenes to derive more accurate Gaussian initialization by enforcing global and local consistency.
Furthermore, we present a geometry-aware pruning strategy that improves optimization stability by retaining high-confidence Gaussians located in non-textured regions.
This scheme facilitates stable updates of the geometric and visual aspects of the scene.
Then, \texttt{LighthouseGS} performs end-to-end optimization of initial camera poses and color inconsistencies via differentiable rasterization, allowing it to correct motion drift and auto-exposure caused by mobile devices.
}

\textcolor{black}{To train and validate our proposed framework, we newly construct real-world and synthetic datasets captured with panorama-style motion.
Our dataset covers various indoor scenes and comprises auto-exposed images with their corresponding camera poses.}
As a result, \methodname{} shows photorealistic rendering quality, outperforming previous neural rendering approaches.
\textcolor{black}{In addition, we further showcase two applications based on the proposed framework: panoramic view synthesis and object placement.}

\indent In summary, our contributions are as follows:
\begin{itemize}
    \item{\textcolor{black}{\methodname{} is a practical NVS framework that allows general users to easily capture indoor scenes in panorama-style motion using a single mobile device.}}
    \item{\textcolor{black}{Based on the indoor planar structure, we introduce a new alignment scheme, plane scaffold assembly, which facilitates initializing 3D Gaussians to fit the scene geometry.}}
    \item{\textcolor{black}{We present a new stable pruning scheme that keeps Gaussians that have high opacity values in textureless regions, enhancing geometric quality and optimization stability.}}
    \item{\textcolor{black}{We introduce geometric and photometric correction strategies to mitigate motion drift and color inconsistencies, resulting in better rendering quality.}}
\end{itemize}

\section{Related work}

\paragraph{Casual Multi-view Capture Motion} \ 
Capturing multi-view images for 3D scene reconstruction has traditionally relied on professional multi-camera rigs~\cite{flynn2019deepview, kaien2020mdp,pozo2019integrated, wilburn2005high,xu2023VRNeRF}. 
However, these systems are often impractical for general users due to their complexity and cost. 
Recent advances have focused on more accessible methods that use a single mobile device for panoramic capture~\cite{bertel2020omniphotos,jang2022egocentric}, multi-view stitching~\cite{hedman2017casual, hedman2018instant}, and view synthesis~\cite{bertel2018megaparallax}.
Panorama-style motion~\cite{sweeney2019structure} has emerged as a popular approach to casual capture, particularly for non-professional users. 
However, this method presents challenges for 3D scene representations, including small baselines, large view changes, and issues with non-textured regions in indoor environments.
\textcolor{black}{To address these challenges, we develop a method tailored to panorama-style inputs for robust indoor 3D scene representations.}

\begin{figure*}[t]
    \centering
    \includegraphics[width=.99\linewidth]{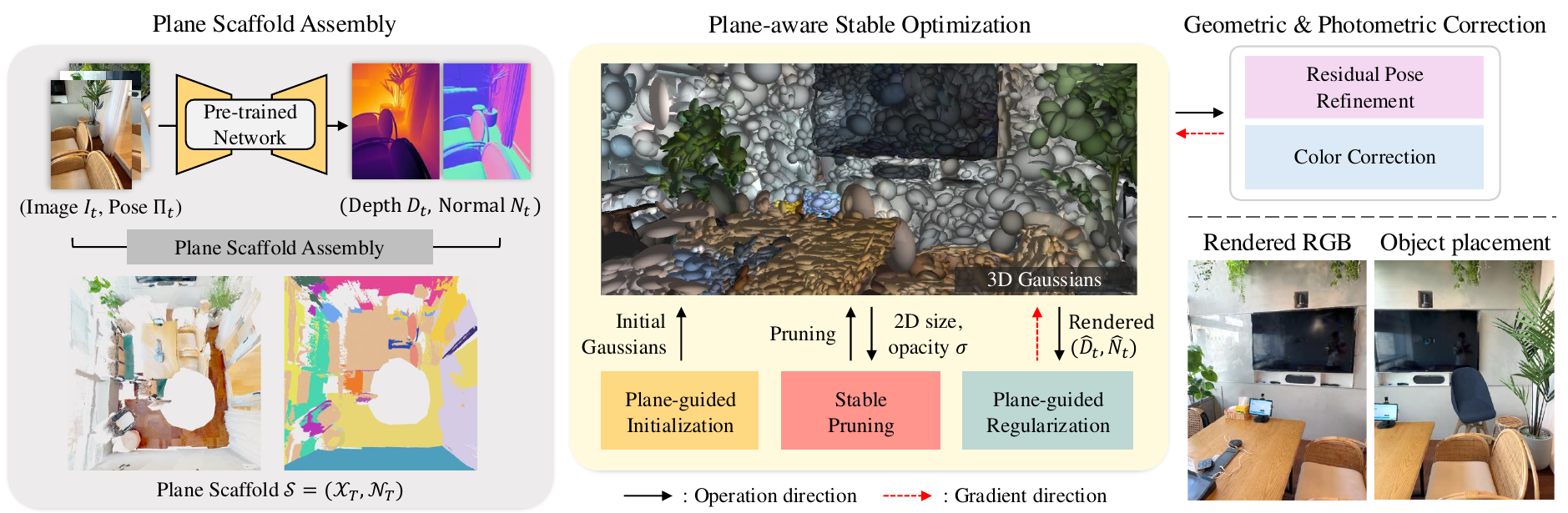}
    % \vspace{-1mm}
    \caption{Overview of \texttt{LighthouseGS}.
    Given consecutive images captured by panorama-style motion with the corresponding rough geometric priors, we construct the plane scaffold that ensures global and local consistency.
    Then, we initialize 3D Gaussians to be aligned to scene geometry and optimize \methodname{} with plane-aware stable optimization.
    To address motion drift and auto-exposure by the use of mobile devices, we additionally correct camera poses and view-dependent colors.
    }
    \label{fig:pipeline}
\end{figure*}
\paragraph{Geometry-aligned 3D Gaussian Splatting}
\textcolor{black}{Recently, 3D Gaussian Splatting (3DGS)~\cite{kerbl20233d} has emerged as a promising framework, offering high-fidelity and real-time view synthesis through explicit 3D Gaussian primitives}
Several works have explored ways to improve the geometric accuracy of 3DGS. Approaches such as 2DGS~\cite{huang20242d} and SuGaR~\cite{guedon2023sugar} have proposed modeling 2D oriented Gaussians, while FSGS~\cite{zhu2025fsgs} and DNGaussian~\cite{li2024dngaussian} incorporate depth guidance in few-view settings. 
\textcolor{black}{GaussianPro~\cite{cheng2024gaussianpro} and DN-Splatter~\cite{turkulainen2025dn} further leverage surface normals alongside depth, reducing ambiguity in non-textured regions.
Beyond these low-level geometric cues, some methods~\cite{li2024geogaussian, ruan2025indoorgs} utilize indoor structural priors, such as lines and planes, to regularize Gaussian optimization.
}
Our work also contributes to this line of research by introducing geometric constraints that consider connectivity between depths and surface normals, tailored for indoor panoramic captures.

\paragraph{Optimizing Camera Pose with Radiance Fields}
Accurate camera poses are crucial for high-quality 3D reconstruction and novel view synthesis. Previous works have explored joint optimization of camera parameters with implicit functions~\cite{lin2021barf,bian2022nopenerf,Chen_2023_dbarf,park2023camp}. 
\textcolor{black}{With the advent of 3DGS, methods like CF-3DGS~\cite{fu2023colmapfree} and HT-3DGS~\cite{ji2025sfm} have emerged to estimate relative poses using depth priors and video frame interpolation, respectively.
Other approaches~\cite{jiang2024construct, meuleman2025fly} enhance camera pose accuracy with correspondence matching, while InstantSplat~\cite{fan2024instantsplat} leverages a trained 3D foundation model~\cite{leroy2024grounding} to regress initial camera pose and dense point maps.
However, these methods struggle to reconstruct a 3D scene under panorama-style motion, which involves large rotation and small translation.
To handle this challenge, we introduce a novel approach that initializes camera poses with AR poses and refines them via residual pose refinement.
}

\section{Lighthouse Gaussian Splatting}
\textcolor{black}{Given a set of images $\{I_t\}$ and initial poses $\{\Pi_t\}$ captured by panorama-style motion, we propose \texttt{\texttt{LighthouseGS}}, a practical 3DGS-based NVS framework for real-time rendering in indoor scenes.
These inputs are easily acquired using smartphone apps~\cite{NeRFCapture} or built-in features like ARKit\footnote{https://developer.apple.com/documentation/arkit} on iOS devices.
As shown in \Fref{fig:pipeline}, we first construct a plane scaffold of aligned 3D points $\mathcal{X}_T$ and their corresponding normals $\mathcal{N}_T$, ensuring global and local consistency (Sec.~\ref{sec4:scaffold}).
From this plane scaffold, we initialize 3D Gaussians to be aligned to the indoor scene geometry, especially in textureless regions.
In the optimization step, \texttt{\texttt{LighthouseGS}} introduces a simple yet effective pruning strategy and then applies geometric constraints to enhance optimization stability in textureless regions of indoor scenes (Sec.~\ref{sec4:optim}). 
While optimizing the 3D Gaussians, we further perform geometric and photometric corrections to resolve the motion drift of input camera poses and color inconsistency caused by auto-exposure (Sec.~\ref{sec4:residual-pose}).
This process allows us to achieve photorealistic rendering from casually captured images without high-end cameras nor SfM.}

\paragraph{Preliminary of 3D Gaussian Splatting}
\textcolor{black}{3DGS~\cite{kerbl20233d} explicitly represents the scene as a set of 3D Gaussian primitives, enabling real-time rendering via differentiable rasterization.
Each Gaussian is expressed as $G(x) = e^{-\frac{1}{2}(x-\mu)^\top\Sigma^{-1}(x-\mu)},$ where $\mu$ indicates 3D mean position and $\Sigma \in \mathbb{R}^{3\times3}$ is a covariance matrix that is decomposed into a rotation matrix~$R\in SO(3)$ and a scale matrix~$S \in \mathbb{R}^{3\times3}$.
}
\textcolor{black}{Concisely, each primitive has learnable parameters $(\mu_i,R_i,S_i,\alpha_i,SH_i)$, where $\alpha$ and $SH$ denote the opacity and spherical harmonics (SH) coefficients.
To optimize these parameters, the covariance matrix is transformed into the camera space through local affine approximation~\cite{zwicker2002ewa}.
Then, during the $\alpha$-blending stage, rendered color $C$ is computed by $C = \sum_{i}^{N}c_i\alpha_iT_i$, where $T_i=\prod_{j=1}^{i-1}\left(1-\alpha_j\right)$ is the transmittance, $c_i$ is the coefficient derived from the SH basis functions, and $N$ is the number of ordered Gaussians overlapping with the pixels.
}

\subsection{Plane Scaffold Assembly}   \label{sec4:scaffold}
\begin{figure}[t]
    \centering
    \includegraphics[width=.99\linewidth]{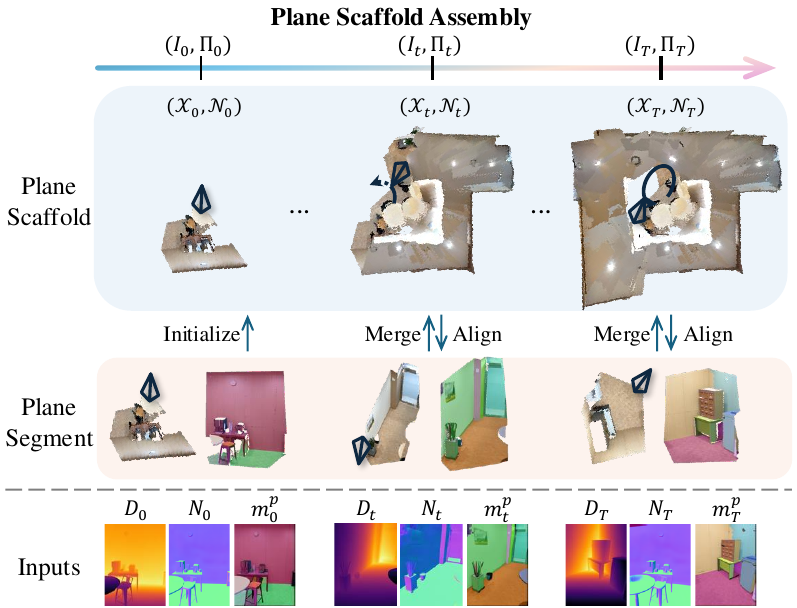}
    % \vspace{-1mm}
    \caption{Overview of plane scaffold assembly.
    \textcolor{black}{Inputs are sequentially merged into the plane scaffold.} 
    The estimated depth from the current frame is globally-to-locally aligned with the projected global points from the previous set. 
    }
    \label{fig:planescaffold}
\end{figure}

We present a new alignment scheme, \emph{plane scaffold assembly}, that constructs globally aligned dense 3D points, leveraging planar structures from the rough geometric priors (\ie, monocular depth estimation~\cite{yang2024depth} and ARKit poses), as shown in~\Fref{fig:planescaffold}.
\textcolor{black}{Although these priors provide fairly accurate information, monocular depths between overlapping views suffer from scale ambiguity, resulting in inaccurate Gaussian initialization.}
\textcolor{black}{To resolve this issue, we propose a two-stage approach for plane scaffold assembly: image-wise global alignment and plane-wise local alignment.}

\paragraph{Image-wise Global Alignment}
Given previous global points {$\mathcal{X}_{t-1}$} and a new depth image $D_t$, image-wise global alignment aims to predict an affine transformation that adjusts a given depth image to fit the global points.
Note that directly aligning depth with global points in 3D is a non-trivial problem due to the requirement of point-wise correspondences.
Instead, we project the global points into the current frame and then align them in 2D space.

\begin{figure}[t]
    \centering
    \includegraphics[width=0.85\linewidth]{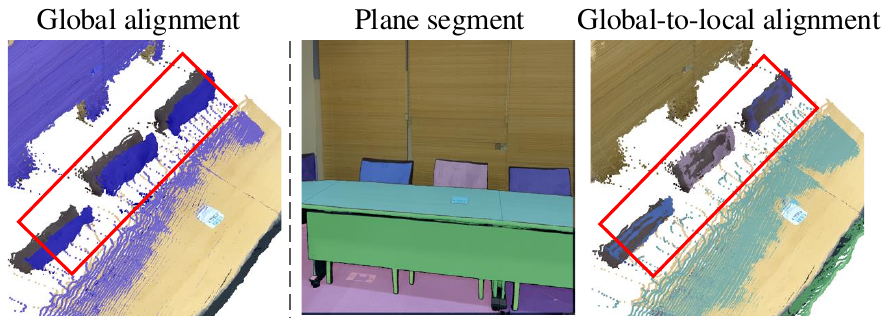}
    % \vspace{-1mm}
    \caption{Effect of plane-wise local alignment. 
    Although the blue points of global alignment include local inconsistency, plane-wise local alignment ensures local consistency.
    }
    \label{fig:planealign}
\end{figure}

We represent an adjusted depth map by affine transformation as $\Bar{D}_t = \alpha_tD_{t} + \beta_t$ with the learnable per-view scale and shift parameters $(\alpha_{t}$, $\beta_{t})$ $\in$ $\mathbb{R}$.
These parameters are optimized via a gradient descent manner to minimize error between the current depth and projected global points, similar to the depth alignment in Text2Room~\cite{hoellein2023text2room}.
While this image-wise global alignment yields overall improvements, relying on a single scale and shift parameter to adjust 3D space for various objects in indoor scenes causes local inconsistency, as shown in \Fref{fig:planealign}.
\textcolor{black}{Note that, as the reference scale, we initialize a set of aligned 3D points $\mathcal{X}_0$ from the first depth frame $D_0$ by back-projection.}
% 
% \textcolor{blue}{
% For this reason, we exploit the scale and shift parameters of the global alignment as the initial values and then refine parameters for each local planar region.
% }

\paragraph{Plane-wise Local Alignment}
We optimize per-plane scale and shift values to enforce local consistency on globally aligned 3D points. 
To this end, we leverage planar geometry in indoor scenes, where depth values with similar normal directions in local regions can be grouped into the same plane.
We extract such planar regions via efficient mean-shift clustering~\cite{Yu2019plane} on an estimated depth $D_t$ and surface normal map $N_t$, even when the exact number of planes is unknown.
We represent each planar region as a plane segment mask {$m_{t}^{p} \in \mathbb{R}^{H\times W}$} as a form of a binary mask.
{Then, we define the per-plane scale and shift parameters $(\gamma_t^p,\delta_t^p)$ for each plane $p$.}
Initial values are initialized with the global parameters $(\alpha_t,\beta_t)$. 
The scaled depth $\Bar{D}_{t}^{p}$ for each plane $p$ is computed by multiplying the plane mask as follows:
\begin{equation}
    {\Bar{D}_{t}^{p} = m^p_t\odot (\gamma_t^p D_t^p+\delta_t^p).}
\end{equation}
These local plane regions are optimized while minimizing the projection loss between each plane segment and the projected depth within the overlapped pixels as follows:
\begin{equation}
    \underset{\gamma_t^p,\delta_t^p}{argmin}||{M}\odot(\Bar{D}_{t}^{p} -  \pi(\mathcal{X}_{t-1};\Pi_t ))||_1,
    \label{eq:local_align}
\end{equation}
where ${M}$ denotes a mask for the overlapped pixels between the scaled plane depth $\Bar{D}_{t}^{p}$ and the projected global points.
Finally, we back-project the adjusted plane segments to obtain the 3D points in the world space.
\textcolor{black}{They are merged to the global points $\mathcal{X}_{t-1}$ except for the overlapping pixels by:}
\begin{equation}
    \mathcal{X}_{t} = \mathcal{X}_{t-1} \cup \pi^{-1}(\Bar{D}_{t}^{p}; \Pi_t),
    \label{eg:merge_pt}
\end{equation}
where $\pi^{-1}(\cdot; \cdot)$ denotes a back-projection function that back-projects \textcolor{black}{globally-to-locally aligned} planes into the world space using the camera pose.
Thanks to the local plane-wise alignment, we observe that the 3D points are correctly aligned across various objects or planar regions.
In addition, each 3D point has a normal direction belonging to its plane, which guides the initialized 3D Gaussians to be aligned well with the scene geometry from the power of plane hints for indoor environments.
{The transformed plane normal directions are aggregated into the plane scaffold as a set of normal vectors $\mathcal{N}_t$ in a similar manner to what is done in \Eref{eg:merge_pt} with the global 3D points $\mathcal{X}_t$.}

Finally, we generate globally-to-locally aligned points with plane information called the plane scaffold $\mathcal{S} = (\mathcal{X}_T,\mathcal{N}_T)$. 
It successfully estimates a set of plane-guided 3D points for initializing 3DGS on our practical and challenging panorama-style motion that fails on COLMAP~\cite{schonberger2016structure}, as shown in \Fref{fig:plane_scaffold_points}.
We downsample this plane scaffold to keep the proper number of points.

\subsection{Plane-aware Stable Optimization}    \label{sec4:optim}
\paragraph{Plane-guided Initialization}
\textcolor{black}{Given a set of aligned 3D points, we initialize 3D Gaussian primitives to be aligned to the surface normals in the plane scaffold.}
Following 3DGS~\cite{kerbl20233d}, a set of Gaussians begins with the position of the global 3D points.
The initial scale of each Gaussian is determined by the average distance from its nearest neighbors, resulting in an isotropic shape.
These isotropic Gaussians roughly populate the 3D world space without leaving empty regions, which does not represent the planar structure largely distributed in indoor scenes.
Thus, we compress the initialized 3D Gaussians by minimizing the scale of the axis closest to the surface normal.
By assigning the minimal value to the scale, the 3D Gaussians are initialized with thin structures flattened along the surface plane.

\begin{figure}[t]
    \centering
    \includegraphics[width=0.99\linewidth]{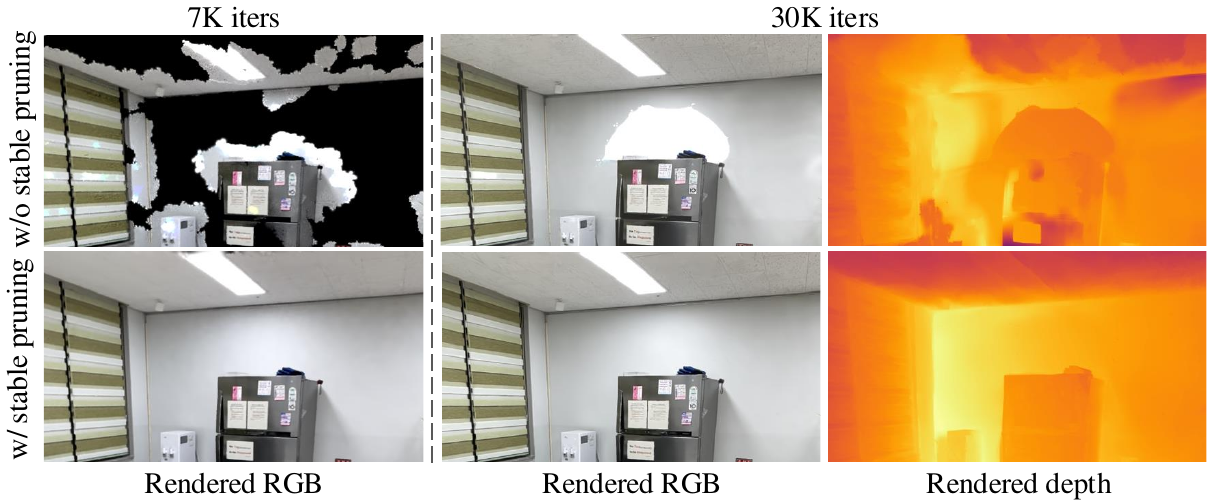}
    \caption{
    Effect of stable pruning.
    \textcolor{black}{Unlike previous pruning schemes, stable pruning keeps highly confident Gaussians, preserving precise scene geometry in textureless areas.}
    }
    \label{fig:stablepruning}
\end{figure}

\paragraph{Stable Pruning}
The existing pruning scheme~\cite{kerbl20233d} often removes oversized Gaussians in non-textured regions. 
This degrades optimization stability since the empty holes are filled inappropriately, causing geometric artifacts after optimization (see top part of \Fref{fig:stablepruning}).

\textcolor{black}{To prevent such artifacts, we introduce a stable pruning strategy that retains reliable 3D Gaussians in textureless regions. 
Intuitively, 3D Gaussians aligned with the surface possess high opacity values.
In addition, since textureless regions do not require fine details, they can be effectively represented with larger Gaussians.
Motivated by these observations, we assess the reliability of 3D Gaussians based on their opacity.}
\textcolor{black}{Specifically, we first collect pruning candidates for oversized Gaussians within the overlapping regions in the image domain.}
Then, we retain the high-confidence Gaussians among the candidates whose opacity value is higher than a threshold $0.5$.
This simple yet effective strategy enhances optimization stability by retaining large Gaussians, thereby preventing large holes in non-textured regions.
\textcolor{black}{As a result, stable pruning produces smooth geometry without floaters in textureless regions, ensuring precise scene representation (see bottom part of \Fref{fig:stablepruning}).}

\paragraph{Plane-guided Regularization}
While the 3D Gaussians are initialized with the proposed plane scaffold, relying solely on the rendering loss does not guarantee their geometric accuracy during optimization. 
Thus, we apply additional loss functions to align the 3D Gaussians with the scene geometry.

\vspace{1mm} \noindent {\textit{\textcolor{black}{Angular Loss.}}} \
To enforce each Gaussian to be aligned with the surface, we constrain its normal direction as follows:
\begin{equation}
    \mathcal{L}_{cos} =1 - cos(\hat{N}, N),
\end{equation}
{where $\hat{N}$ is the rendered normal map, and $N$ is the estimated normal map from the pre-trained network~\cite{bae2024dsine}.
}
Following~\cite{cheng2024gaussianpro}, we can compute $\hat{N}_t$ by replacing each Gaussian's color $c_{i}$ with its normal $n_{i}$, where $n_{i}$ is a rotation axis having the minimal scale value.

\vspace{1mm} \noindent \textit{\textcolor{black}{Flatten Loss.}} \
We also regularize the minimal scale of each Gaussian to be flattened along the surface, inspired by NeuSG~\cite{chen2023neusg}:
\begin{equation}
    \mathcal{L}_{flat} =||min(s_1,s_2,s_3)||_1,
\end{equation}
where \textcolor{black}{$s_i$ denotes the scale value of each axis of the 3D Gaussian in the world space.}

\vspace{1mm} \noindent {\textit{\textcolor{black}{Normal Smoothness Loss.}}}
3DGS often struggles with representing smooth geometry due to its unstructured form.
To mitigate this issue, we apply a total variation term~\cite{rudin1992nonlinear} to the rendered normal:
\begin{equation}
\label{eq:normal_smooth}
    \mathcal{L}_{smooth} = \frac{1}{L}\sum_{i,j} |\nabla_x\hat{N}(i,j)| + |\nabla_y\hat{N}(i,j)|,
\end{equation}
\textcolor{black}{where $(i, j)$ denotes the pixel coordinates and $\nabla_x \hat{N}$ and $\nabla_y \hat{N}$ are the horizontal and vertical gradient of the rendered normal map $\hat{N}$, and $L$ is the number of pixels.}

\vspace{1mm} \noindent {\textit{Depth-to-Normal Consistency Loss.}} \
We regularize the rendered depth map to be consistent in local regions using the normal map.
\textcolor{black}{To do this, we back-project the rendered depth map $\hat{D}$ into a per-pixel 3D location map.
Then, depth-to-normal consistency is enforced by encouraging the horizontal and vertical gradients of 3D locations to be orthogonal to the corresponding normal direction:}
\begin{equation}
    \mathcal{L}_{d2n} = \frac{1}{L} \sum_{i,j} | \nabla_x \hat{D}(i,j) \cdot N(i,j) | + | \nabla_y \hat{D}(i,j) \cdot N(i,j) |,
\end{equation}
where $\nabla_x \hat{D}$ and $\nabla_y \hat{D}$ are the spatial gradient of the 3D location $\hat{D}$ and $L$ is the number of pixels.

\subsection{Geometric and Photometric Correction}
\label{sec4:residual-pose}
Plane scaffold assembly and plane-aware stable optimization mitigate the limitations of existing 3DGS in indoor scenes.
\textcolor{black}{However, our framework still inherits challenges like motion drift and auto-exposure/white balance from mobile capture.}
To alleviate these issues, we present two strategies: residual pose refinement and color correction.

\paragraph{Residual Pose Refinement}
Inspired by PoRF~\cite{bian2024porf}, we optimize residual pose instead of directly refining camera pose parameters.
\textcolor{black}{
We initialize residual pose parameters for each frame as an identity rotation matrix (quaternion form) and zero translation vector. 
}
During optimizing 3DGS, a quaternion and translation vector are transformed into the residual pose $\Delta\Pi \in SE(3)$. 
To obtain an adjusted pose, the residual pose is multiplied by the initial pose:
\begin{equation} \label{eq:residual_pose}
    \Tilde{\Pi}_t = \Delta\Pi_t \cdot \Pi_t,
\end{equation}
where $\Tilde{\Pi}_t$ indicates the adjusted pose with the learnable residual parameters, {but the camera intrinsic is fixed.}
Then, we compare the rendered image utilizing the adjusted pose with the ground truth image.
Since the image is rendered with the adjusted pose in a differentiable renderer, gradients can be backpropagated to update the residual pose.
With the residual pose refinement, we render geometrically consistent views while alleviating motion drift, particularly in later sequences where drift errors accumulate.

\paragraph{Color Correction}
Captured images from mobile devices are typically subject to auto-exposure and white balance, leading to color inconsistencies across different viewpoints.
Color variations at the same location disrupt the consistency of 3DGS in learning geometry and color.
Varying colors across different views are corrected via a simple color transformation strategy parameterized by learnable white balancing $w \in \mathbb{R}^3$ and brightness $b \in \mathbb{R}^3$ as channel-wise parameters.
\textcolor{black}{A rendered image $\hat{I}_t$ is transformed into the color corrected image $\Tilde{I}_t$ before calculating rendering loss:}
\begin{equation}
    \Tilde{I}_t = w_t\hat{I}_t+b_t.
\end{equation}
The learnable coefficients are updated via differentiable rasterization, backpropagating the gradients into these parameters.
\textcolor{black}{This approach compensates for color differences and learns a global color space to improve visual quality.}

\subsection{Objective Function}
\textcolor{black}{To train the proposed \texttt{\texttt{LighthouseGS}}, the final loss term $\mathcal{L}$ consists of a photometric and geometric constraint:
\begin{equation}
    \mathcal{L} = \mathcal{L}_{color} + \mathcal{L}_{geo}.
\end{equation}
Photometric loss $\mathcal{L}_{color}$ guides explicit 3D Gaussians for reducing photometric differences between ground truth images $I$ and corrected images $\Tilde{I}$ from rendered image $\hat{I}$:}
\begin{equation}
    \mathcal{L}_{color} = \lambda_{l1}\mathcal{L}_{l1} + \lambda_{D-SSIM}\mathcal{L}_{D-SSIM},
\end{equation} 
\begin{equation}
   \mathcal{L}_{l1} = ||I - \Tilde{I}||_1, \   \mathcal{L}_{D-SSIM} = 1 - SSIM(I,\Tilde{I}),
\end{equation} 
\textcolor{black}{where $\lambda_{l1}$ and $\lambda_{D-SSIM}$ are set to $0.8$ and $0.2$, respectively.}

Geometric loss $\mathcal{L}_{geo}$ regularizes the rendered geometry to align 3D Gaussians along the surface using prior geometry.
\textcolor{black}{The total geometric constraints are given by:}
\begin{equation}
    \mathcal{L}_{geo} = \lambda_{normal}(\mathcal{L}_{cos} + \mathcal{L}_{flat} + \mathcal{L}_{smooth}) + \lambda_{d2n}\mathcal{L}_{d2n},
\end{equation}
\textcolor{black}{where we set $\lambda_{normal} = 0.05$ and $\lambda_{d2n} = 0.2$.}
\begin{figure*}[t]
    \centering
    \includegraphics[width=0.99\linewidth]{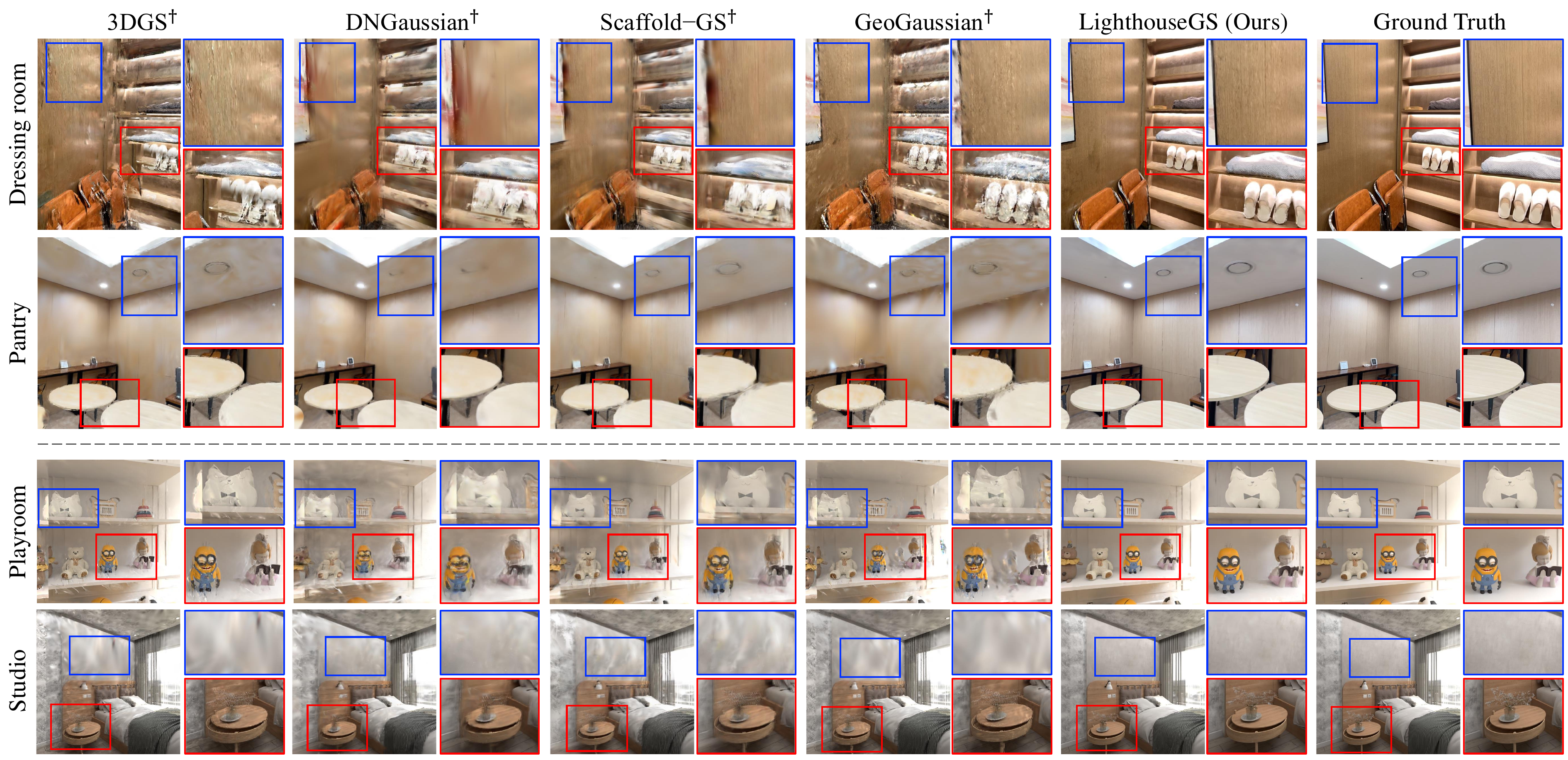}
   \caption{
   \textcolor{black}{
    Qualitative comparisons on the real-world (top) and synthetic datasets (bottom).
    In contrast to other methods, \texttt{LighthouseGS} clearly reconstructs details of textured shapes without blurry artifacts (see red boxes). 
    Also, we can observe that our method preserves accurate scene geometry without floaters in non-textured regions (see blue boxes).}
    }
    \label{fig:qualitative_real}
\end{figure*}

\section{\textcolor{black}{Experiments}}
\subsection{Dataset} \label{sec4:dataset}
Due to the absence of indoor scene datasets captured by panorama-style motion, we construct a new dataset in real and synthetic environments to evaluate our \methodname{} framework.
\textcolor{black}{We visualize the collected dataset in the supplementary material.}

\paragraph{Real-world Scenario}
We use Nerfcapture~\cite{NeRFCapture} that saves a set of images with the camera parameters from iPhone's ARKit.
We rotate the device with outstretched half-arms while capturing the region of interest. 
A set of images covers almost the entire space from the motion center.
This natural motion does not strictly follow the original spherical motion~\cite{ventura2016structure}, and panorama-style motion means rather a more practical and less constrained method.
\textcolor{black}{Our dataset comprises five different indoor environments, each with up to 100 images at $1920 \times 1440$ resolution.}

\paragraph{Synthetic Scenario}
\textcolor{black}{We create a synthetic dataset with Blender \cite{blender}, which supports physically-based rendering.}
We render synthetic images at a radius of 20 or 30 cm from the fixed center of panorama-style motion, depending on the scale of the scene.
Based on a statistical analysis of ARKit tracking~\cite{kim2022benchmark}, we further add drift noise to the ground truth camera poses to mimic pose errors.
\textcolor{black}{It consists of five indoor scenes, each with 100 auto-exposure images at $1024 \times 1024$ resolution and their corresponding camera poses.}

\subsection{Implementation Details} \label{sec4:implementation_details}
\methodname{} is built on the gsplat~\cite{ye2025gsplat}.
\textcolor{black}{For Gaussian densification, a gradient threshold is set to 0.0008 and 0.0004 for real and synthetic scenes, respectively.} 
All other configurations follow the original settings in 3DGS~\cite{kerbl20233d}, and the experiments are conducted on a single RTX 4090 GPU.
Following \cite{fu2023colmapfree}, we freeze the trained 3D Gaussians and optimize the camera poses and tone mapping parameters of test views over 20K iterations to evaluate unseen viewpoints.
\textcolor{black}{Then, we measure PSNR, SSIM, and LPIPS.} 

\subsection{Evaluation} \label{sec4:eval}
\textcolor{black}{We evaluate \texttt{LighthouseGS} with state-of-the-art 3DGS-based approaches: 3DGS~\cite{kerbl20233d}, DNGaussian~\cite{li2024dngaussian}, Scaffold-GS~\cite{scaffoldgs}, and GeoGaussian~\cite{li2024geogaussian}.
All comparisons rely on pre-computed 3D points and camera poses from COLMAP~\cite{schonberger2016structure}. 
However, it fails under panorama-style motion, making it impossible to execute the methods.
To address this, we initialize baselines using 3D points from our proposed plane scaffold assembly and ARKit poses for a fair comparison (denoted as $\dagger$).}

\begin{table}[t]
\centering
\small
\resizebox{0.9\linewidth}{!}{%
\begin{tabular}{llccc}
\toprule
Dataset & Method & PSNR $\uparrow$ & SSIM $\uparrow$  & LPIPS $\downarrow$ \\
\midrule
\multirow{5}{*}{Real-world}
  & 3DGS$^\dagger$        & 20.60 & 0.684 & 0.387 \\
  & DNGaussian$^\dagger$  & 21.10 & \underline{0.724} & 0.394 \\
  & Scaffold-GS$^\dagger$ & \underline{21.11} & 0.707 & 0.338 \\
  & GeoGaussian$^\dagger$ & 20.53 & 0.687 & \textbf{0.311} \\
  & LighthouseGS & \textbf{25.06} & \textbf{0.809} & \underline{0.317} \\
\midrule
\multirow{5}{*}{Synthetic}
  & 3DGS$^\dagger$        & 24.10 & 0.790 & 0.252 \\
  & DNGaussian$^\dagger$  & 23.80 & \underline{0.806} & 0.312 \\
  & Scaffold-GS$^\dagger$ & 23.64 & 0.781 & \underline{0.251} \\
  & GeoGaussian$^\dagger$ & \underline{24.14} & 0.800 & 0.263 \\
  & LighthouseGS & \textbf{28.86} & \textbf{0.898} & \textbf{0.122} \\
\bottomrule
\end{tabular}
}
% \vspace{-1mm}
\caption{Quantitative comparisons on real and synthetic datasets.
$\dagger$ indicates methods that initialize with plane scaffold assembly. 
\textcolor{black}{We highlight the best and second performances in bold and underline.}
}
\label{tab:quantitative}
\end{table}

\paragraph{Quantitative Evaluation}
\textcolor{black}{We report average quantitative results for the collected real-world and synthetic datasets (see \Tref{tab:quantitative}).}
\texttt{LighthouseGS} outperforms previous methods across all metrics on both datasets.
These significant improvements demonstrate that \texttt{LighthouseGS} effectively resolves inherent problems posed by capturing indoor scenes with panorama-style motion using mobile devices.
In other words, our plane-aware optimization scheme handles artifacts in textureless regions while pose refinement mitigates motion drift, resulting in more accurate reconstructions.
It should be noted that per-scene metrics and \textcolor{black}{computational efficiency (\emph{e.g.}, memory usage, training time, and FPS)} are provided in the supplementary material.

\paragraph{Qualitative Evaluation}
Figure~\ref{fig:qualitative_real} shows novel view synthesis results on a real-world and synthetic dataset.
Overall, \texttt{LighthouseGS} achieves high-fidelity rendering while preserving accurate scene geometry.
\textcolor{black}{For example, unlike comparison methods that cause floaters and over-smoothed surfaces in textureless regions, \texttt{LighthouseGS} suffers less from such artifacts (see blue boxes).
We deduce that retaining high-confidence 3D Gaussians in these areas by stable pruning prevents artifacts, resulting in accurate scene geometry.}
Moreover, our method is robust to inaccurate input camera poses and auto-exposure images (see red boxes). 
As illustrated at the bottom of \Fref{fig:qualitative_real}, \texttt{LighthouseGS} yields similar performance on the synthetic dataset.
In particular, due to changing exposure values as the camera rotates away from the window where the light comes in, other methods render visually inconsistent results (see blue boxes in the playroom).
In contrast, \texttt{LighthouseGS} compensates for color differences by learning global color space through color correction, which synergizes with stable optimization and pose refinement to improve rendering quality.
\begin{table}[t]
    \centering
    \small
    \resizebox{0.95\linewidth}{!}{%
    \begin{tabular}{l|cccc}
        \toprule
        
        Methods      & PSNR $\uparrow$ & SSIM $\uparrow$ & LPIPS $\downarrow$ & \# of Gaussians $\downarrow$ \\
        \midrule
        3DGS$^\dagger$ &  22.12 & 0.803 & 0.335 & 3.9M \\
        w/o res. pose & 23.25 & 0.850 & 0.339 & \textbf{0.46M} \\
        w/o color corr.  & 24.64 & \underline{0.884} & 0.298 & 0.6M \\
        w/o stable optim. & \underline{26.47} & \underline{0.884} & \textbf{0.284} & 0.92M \\
        Ours (full model) & \textbf{26.80} & \textbf{0.888} & \underline{0.287} & \underline{0.51M} \\
        \bottomrule
    \end{tabular}
    }
    % \vspace{-1mm}
    \caption{\label{tab:method_ablation}
        Module ablation. We ablate our three important modules in \textit{pantry}. 
        The best scores are highlighted as bold. 
    }
\end{table}
\begin{figure}[t!]
    \centering
    \includegraphics[width=0.82\linewidth]{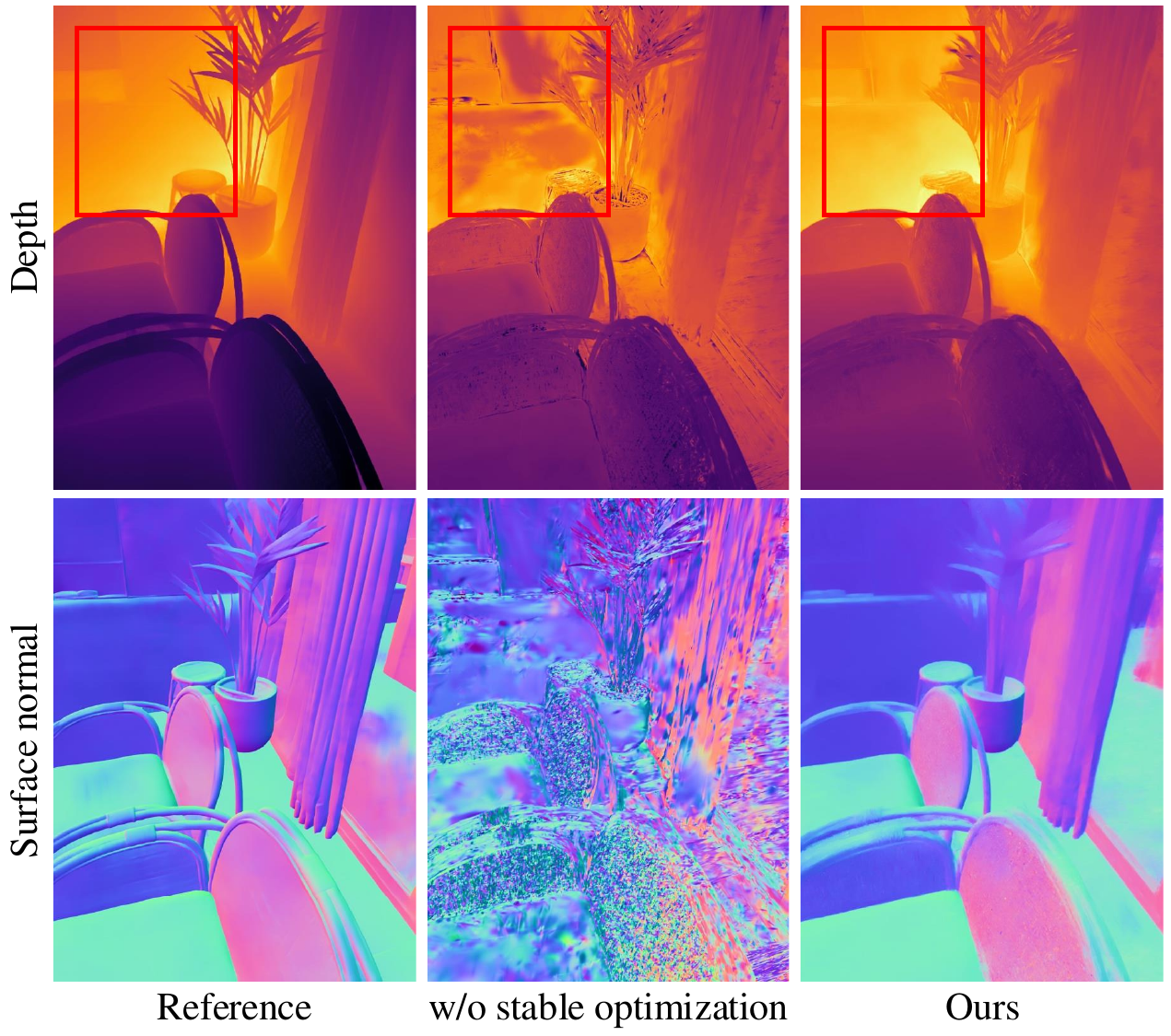}
    \caption{Effect of stable optimization.
    \textcolor{black}{Stable optimization ensures Gaussian primitives can render consistent geometry, alleviating textureless artifacts and floaters.}
    }
    \label{fig:ablate_stable_optim}
\end{figure}
\textcolor{black}{
As shown in \Fref{fig:ablate_stable_optim}, our method well represents soft furnishings and curved objects (\emph{e.g.}, curtains and vase), demonstrating generalization beyond planar structures.}
\begin{table}[t]
    \centering
    \small
    \resizebox{0.85\linewidth}{!}{%
    \begin{tabular}{l|ccc}
        \toprule
        Methods      & PSNR $\uparrow$ & SSIM $\uparrow$ & LPIPS $\downarrow$ \\
        % Methods      & PSNR & SSIM & LPIPS \\
        \midrule
        COLMAP & N/A & N/A & N/A \\
        Spherical SfM & 17.41 & \underline{0.660} & \underline{0.440} \\
        \textcolor{black}{ARKit + COLMAP}  & \underline{17.72} & 0.654 & 0.443 \\
        ARKit + Plane Scaffold (Ours) & \textbf{23.52} & \textbf{0.789} & \textbf{0.300} \\
        \bottomrule
    \end{tabular}
    }
    % \vspace{-1mm}
    \caption{\label{tab:init_ablation}
    Initialization ablation. \textcolor{black}{We compare different initializations with native 3DGS in the conference room}.
    }
\end{table}
\subsection{Ablation Study} \label{sec4:ablation}
\paragraph{Module Ablation}
We ablate each proposed component of our framework to demonstrate its effectiveness compared to the baseline (see \Tref{tab:method_ablation}).
Each module contributes to improved performance across all metrics, even with fewer Gaussians.
\textcolor{black}{Since we follow adaptive density control, which relies on view space gradients, the residual pose refinement and color correction enhance rendering quality, resulting in smaller gradients that reduce the number of Gaussians.}
In addition, we observe that stable optimization is effective in further improving performance.
\textcolor{black}{
Although stable optimization marginally improves photometric quality, it leads to a significant reduction in the number of Gaussians and enhances scene geometry fitting, as shown in~\Fref{fig:ablate_stable_optim}.
}
We deduce that this strategy prevents mis-densification in non-textured regions by preserving highly confident 3D Gaussians.
\textcolor{black}{
To justify the proposed modules, we include further ablation studies in the supplementary materials.
}

\begin{figure}[t]
    \centering
    \includegraphics[width=0.95\linewidth]{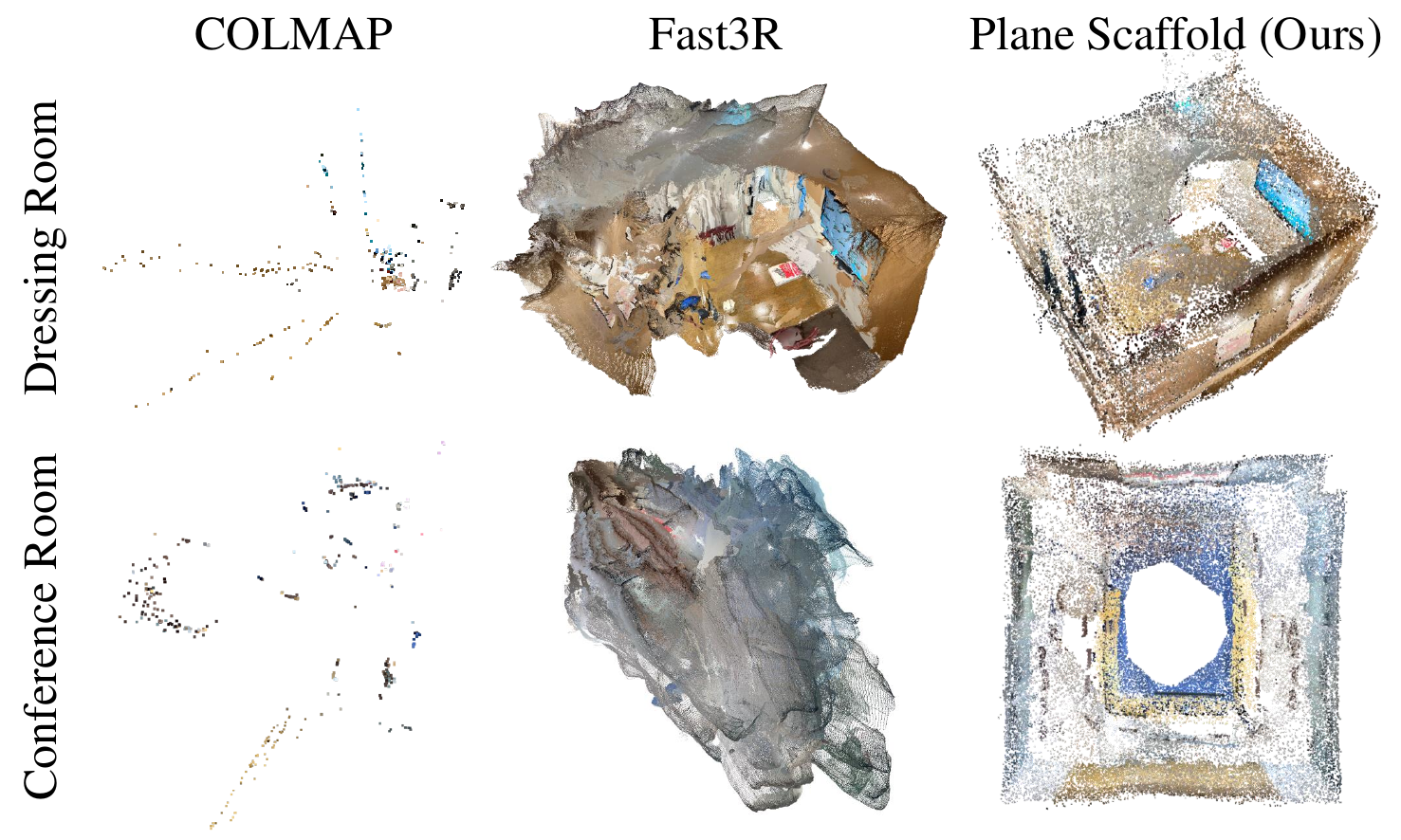}
    % \vspace{-2mm}
    \caption{Qualitative comparisons of initial point clouds.}
    \label{fig:plane_scaffold_points}
\end{figure}

\paragraph{Initialization Ablation}
In \Tref{tab:init_ablation}, we validate the efficacy of our plane scaffold assembly by altering the initialization algorithm in 3DGS.
Since COLMAP often fails under a panorama-style motion setting, we triangulate a bundle of points with ARKit poses (denoted as ARKit + COLMAP).
\textcolor{black}{While ARKit enables COLMAP reconstruction, it still produces inaccurate camera poses and 3D points, resulting in lower rendering quality.}
Spherical SfM~\cite{ventura2016structure} is limited due to the strong constraint that cameras must lie on the sphere to operate successfully.
\textcolor{black}{Also, \Fref{fig:plane_scaffold_points} shows initial point clouds generated by various methods, including the geometric foundation model Fast3R~\cite{fast3r}.}
As a result, our plane scaffold assembly shows robust performance in both qualitative and quantitative aspects.

\begin{table}[t!]
    \centering
    \small
    \resizebox{0.8\linewidth}{!}{
    \begin{tabular}{l|ccc}
        \toprule
        Methods       & PSNR $\uparrow$ & SSIM $\uparrow$ & LPIPS $\downarrow$ \\
        \midrule
        ZoeDepth+DSINE    		 & 26.25 & 0.839 & \underline{0.194}  \\
        DA-V2+Omnidata    		 & \underline{26.87} & \underline{0.841} & \textbf{0.188}  \\
        DA-V2+DSINE (Ours)          & \textbf{26.96}		
 & \textbf{0.853} & 0.219 \\
        \bottomrule
    \end{tabular}
    }
    \caption{\textcolor{black}{Ablation on monocular priors. We report the sensitivity of monocular priors by alternating different backbone networks.
    }
    }
    \label{tab:monodepth}
\end{table}
\noindent\textbf{Sensitivity to Priors.}
\textcolor{black}{
We utilize monocular depth and normal estimation to construct globally-to-locally aligned 3D points in the plane scaffold assembly. 
To assess sensitivity to rendering quality, we conduct quantitative experiments by replacing Depth Anything V2 (DA-V2)~\cite{yang2024depth} and DSINE~\cite{bae2024dsine} with ZoeDepth~\cite{bhat2023zoedepth} and Omnidata~\cite{eftekhar2021omnidata}, respectively.
As reported in \Tref{tab:monodepth}, even with lower quality depth or normal inputs, our method shows comparable performance, demonstrating its robustness to monocular priors.
Moreover, with ongoing advances in monocular estimation, such sensitivity will become less critical in practice. 
}
\section{Applications}
\paragraph{Object Placement}
\textcolor{black}{We demonstrate the applicability of \texttt{LighthouseGS} to AR applications such as object placement (see \Fref{fig:pipeline}). 
Accurate placement of virtual objects at desired locations requires precise scene geometry to ensure seamless alignment with the physical environment. 
With the geometrically aligned scene from \texttt{LighthouseGS}, virtual objects can be naturally inserted into the scene.}

\paragraph{Panoramic View Synthesis}
Panoramic imaging presents the entire indoor scene in a single image, making it useful in AR/VR applications.
Thanks to our panorama-style motion setting, we can render panoramic images from an unseen viewpoint with the sphere-based rasterizer~\cite{huang2024error}.
\textcolor{black}{Additional results, including object placement and panoramic rendering, are available in the supplementary material.}
\section{Conclusion} \label{sec:conclusion}
We propose \texttt{LighthouseGS}, a practical 3DGS-based novel view synthesis framework for indoor scenes from panorama-style motion capture with a single mobile device.
By incorporating the planar structure into the entire pipeline, we introduce a planar scaffold for consistent 3D alignment and a plane-aware optimization strategy that retains highly confident Gaussians in non-textured regions.
We further correct inaccurate camera poses and auto-exposure artifacts to enhance visual quality. 
\textcolor{black}{Experiments demonstrate photorealistic rendering with accurate scene geometry, broadening the applicability of 3DGS to casual mobile capture scenarios.}

\section*{Acknowledgements}
This work was supported by Bucketplace, by Institute of Information \& communications Technology Planning \& Evaluation (IITP) grant funded by the Korea government (MSIT) (No.RS-2022-II220907, Development of AI Bots Collaboration Platform and Self-organizing; No.RS-2020-II201336, Artificial Intelligence Graduate School Program (UNIST)) and by the InnoCORE program of the Ministry of Science and ICT (25-InnoCORE-01).

{
    \small
    \bibliographystyle{ieeenat_fullname}
    \bibliography{main}
}

\end{document}